\documentclass[12pt]{article}
\usepackage{hyperref}
\usepackage{amsmath}
\usepackage{graphicx}
\usepackage{amssymb} % Symbols for itemise
 \usepackage{xcolor}
\definecolor{myblue}{rgb}{0.14,0.11,0.49}
\definecolor{myred}{rgb}{0.74,0.22,0.15}
\definecolor{mygreen}{rgb}{0.05,0.52,0.42}
\definecolor{myyellow}{rgb}{0.96,0.92,0.13}
\definecolor{myorange}{rgb}{1,0.61,0.36}
\definecolor{mypurple}{rgb}{0.71,0.02,1}
\newcommand{\Mat}[1]{{{\boldsymbol{#1}}}}
\newcommand{\abs}[1]{\left\vert#1\right\vert}
\def\be{\begin{equation}}
\def\ee{\end{equation}}
\def\bea{\begin{eqnarray}}
\def\eea{\end{eqnarray}}
\def\bi{\begin{itemize}}
\def\ei{\end{itemize}}
\def\noi{\noindent}
\def\dd{\mathrm{d}}

\date{}

\title{{\bf Comment on \\``Spin in an arbitrary gravitational field"}}

\author{Mayeul Arminjon\\
\small\it Laboratory ``Soils, Solids, Structures, Risks'', 3SR\\ \small\it (CNRS and Universit\'es de Grenoble: UJF, Grenoble-INP),\\\small\it BP 53, F-38041 Grenoble cedex 9, France.}

\begin{document}
\maketitle
%\vspace{7mm}

\begin{abstract} 
\noi The authors of that work [Phys. Rev. D {\bf 88}, 084014 (2013)]
%[arXiv:1308.4552 (gr-qc)] 
derive quantum-mechanical equations valid for the covariant Dirac equation by restricting the choice of the tetrad field through the use of the ``Schwinger gauge". Yet it has been shown previously that this gauge leaves space for a physical ambiguity of the Hamiltonian operator. It is shown here precisely how this ambiguity occurs with their settings. There is another ambiguity in the Foldy-Wouthuysen Hamiltonian, for the time-dependent case which is relevant here. However, their equations of motion for classical spinning particles are unambiguous.

%\vspace{35mm} 

\end{abstract} 
\vspace{5mm}

\noi Obukhov, Silenko \& Teryaev \cite{OST2013} consider the (standard form of the) covariant Dirac equation in an arbitrary coordinate system in a general spacetime. They derive quantum-mechanical equations and compare them with classical equations. Their first step (i) consists in restricting the choice of the tetrad field by using the ``Schwinger gauge". In a second step (ii), by writing the corresponding expression of the covariant Dirac equation in the Schr\"odinger form while reexpressing the wave function with a specific non-unitary transformation, they get a Hermitian Hamiltonian $\mathcal{H}$. In a third step (iii), they transform it into a new Hamiltonian $\mathcal{H}_\mathrm{FW}$ by using a Foldy-Wouthuysen transformation; based on  $\mathcal{H}_\mathrm{FW}$, they compute the time derivative of a polarization operator. Then (iv) they take the semi-classical limit, which (v) they compare with the Hamiltonian and the equations of motion got for a classical spinning particle.\\

However, in Refs. \cite{A47,A50}, it has been shown that a whole functional space of different tetrad fields still remains available in the Schwinger gauge, and that the Hermitian Hamiltonians deduced from any two of them by precisely the same non-unitary transformation of the wave function as the one used in Ref. \cite{OST2013} are in general physically inequivalent. This means that step (i) is not unique and leads to a physical non-uniqueness at step (ii). It will be shown how exactly this applies. Hence, the subsequent steps of the work \cite{OST2013} are {\it a priori} non-unique, too. This will be discussed also.

\paragraph{1. Non-uniqueness of a tetrad field in the Schwinger gauge. ---}\label{NU-Tetrad} With the parameterization of the spacetime metric used by the authors of Ref. \cite{OST2013} (hereafter OST for short), Eq. (2.1), it appears in the following way.
\footnote{\ 
All equation numbers of the form $(m.n)$ refer to the corresponding equations in Ref. \cite{OST2013}. 
}
As noted by OST, ``the line element (2.1) is invariant under redefinitions
$W^{\widehat a}{}_b\longrightarrow L^{\widehat a}{}_{\widehat c}\,W^{\widehat c}{}_b$ using arbitrary local rotations $L^{\widehat a}{}_{\widehat c}(t,x)\in SO(3)$." [Clearly, here $x$ denotes the triplet $x\equiv (x^a)\ (a=1,2,3)$. Note that OST consider an arbitrary spacetime coordinate system, but do not envisage its change.] Under such a redefinition, the cotetrad field defined by Eq. (2.2): $\theta ^\alpha \equiv  e^\alpha _{\ \, i }\, dx^i \ (\alpha,i =0,...,3)$ changes to $\theta'^\alpha \equiv e'^\alpha_{\ \ i }\, dx^i$ [$(dx^i)$ being the basis of one-forms dual to the natural basis $(\partial _i)$ of the spacetime coordinate system $(x^i)$, where $x^0\equiv t$ according to OST's convention], with 
\be\label{theta'}
e'^{\widehat 0}_{\ \ i }\equiv e^{\widehat 0}_{\ \, i }\ \ (\theta'^{\widehat 0}\equiv \theta ^{\widehat 0}), \quad e'^{\widehat{a}}_{\ \ i} \equiv W'^{\widehat a}{}_b\left(\delta^b_i - cK^b\,\delta^{\,0}_i\right),
\ee
and where
\be\label{W'}
W'^{\widehat a}{}_b \equiv L^{\widehat a}{}_{\widehat c}\,W^{\widehat c}{}_b.
\ee
The tetrad field $u_\alpha \equiv e^i_{\ \, \alpha }\partial _ i$ defined by Eq. (2.3) changes to $u'_\alpha \equiv  e'^i_{\ \, \alpha } \partial _i$, with 
\be\label{u'}
e'^i_{\ \, \widehat 0 }\equiv e^i_{\ \, \widehat 0 }\ \ (u'_{\widehat 0} \equiv u_{\widehat 0}), \quad e'^i_{\ \, \widehat a }\equiv \delta ^i_b\, W'^b_{\ \ \widehat a },
\ee
where 
\be
W'^b_{\ \ \widehat a } \equiv W^b_{\ \, \widehat c }\,P^{\widehat c}_{\ \, \widehat a },
\ee
with $P=(P^{\widehat c}_{\ \, \widehat a }) \ (c, a =1,2,3)$ the inverse matrix of the matrix $(L^{\widehat a}{}_{\widehat c})$: $P=P(t,x)\in SO(3)$, hence $P^{\widehat c}_{\ \, \widehat a }=L^{\widehat a}{}_{\widehat c}$. Thus, the new tetrad field in the Schwinger gauge, Eq. (\ref{u'}), is deduced from the first one by a local Lorentz transformation $\Lambda =\Lambda (t,x)\in SO(1,3)$:
\be\label{Lambda}
u'_\beta =\Lambda ^\alpha _{\ \,\beta }\,u_\alpha , \qquad  \Lambda =\begin{pmatrix} 
1 & 0  & 0 & 0\\
0  &  &  & \\
0 &  & P &\\
0 &   &  &
\end{pmatrix}.
\ee 
Such a redefinition, envisaged by OST themselves, affects many relevant quantities. E.g. in (2.6), $\mathcal{F}^a_{\ \,b}$ becomes
\be\label{F'}
\mathcal{F}'^a_{\ \ b} = \mathcal{F}^a_{\ \,c} P^{\widehat c}_{\ \, \widehat b }.
\ee 
It affects also ${\mathcal Q}_{\widehat{a}\widehat{b}}$ in (2.11), ${\mathcal C}_{\widehat{a}\widehat{b}}{}^{\widehat{c}}$ in (2.12), hence also $\Gamma _{i\alpha \beta }$ in (2.9), (2.10) and (2.8), etc. Hence, {\it a priori,} every result in the ``quantum" part of the paper (sections II-III), as well as every comparison between quantum and classical results (section IV), may depend on the admissible choice of the field of the $3\times 3$ real matrix $W\equiv (W^{\widehat c}{}_b)$ --- that field being determined by the data of the spacetime metric only up to a spacetime dependent rotation matrix $(L^{\widehat a}{}_{\widehat c}(t,x))\in SO(3)$, Eq. (\ref{W'}).

\paragraph{2. Non-uniqueness of the Hermitian Hamiltonian operator (2.15). ---}\label{NU-H} As OST note, the non-unitary transformation (2.14) ``also appears in the framework of the pseudo-Hermitian quantum mechanics" as it is used in Ref. \cite{GorbatenkoNeznamov2011}. More precisely, since the usual relation $g_{\alpha\beta}\, e^\alpha_{\ \,i}\, e^\beta_{\ \,j} = g_{ij}$ mentioned (with a misprint) by OST is equivalent to $g^{\alpha\beta}\, e^i_{\ \,\alpha}\, e^j_{\ \,\beta} = g^{ij}$, the ``Schwinger gauge" condition $e^0_{\ \,\widehat a}=0$ gives 
\be\label{a^0_0}
g^{\widehat 0 \widehat 0}\,(e^0_{\ \,\widehat 0})^2 =g^{00},
\ee 
thus with OST's convention $x^0=t$: $e^0_{\ \,\widehat 0}=c \sqrt{g^{00}}=1/V$ [Eqs. (2.3) and (4.33)]. Instead, with $x^0=ct$, (\ref{a^0_0}) implies $\abs{e^0_{\ \,\widehat 0}}=\sqrt{\abs{g^{00}}}$ independently of the signature \cite{A50,GorbatenkoNeznamov2011}. Therefore, the transformation (2.14):
\be\label{T=a^{-1}I}
\psi =\left(\sqrt{-g}\,e^0_{\ \,\widehat 0}\right)^\frac{1}{2}\Psi 
\ee
is exactly the one used by Gorbatenko \& Neznamov \cite{GorbatenkoNeznamov2011} to transform the Hamiltonian which they note $\widetilde{H}$,  got with some Schwinger tetrad, into the Hermitian Hamiltonian noted $\mathrm{H}_\eta$ by them, Eqs. (67) and (72) in Ref. \cite{GorbatenkoNeznamov2011}. It is also the particular case of the local similarity transformation $T$ in Eq. (18) of Ref. \cite{A50}, corresponding with $S={\bf 1}_4$ (one starts from a Schwinger tetrad). A crucial property of the transformation (\ref{T=a^{-1}I}), that it brings the Hilbert-space scalar product to the ``flat" form \cite{A50,GorbatenkoNeznamov2011}:
\footnote{\
In Eq. (\ref{scalar product})$_1$, $\gamma ^{\widehat 0}$ is the ``$\alpha =0$" constant Dirac matrix (assumed to be ``hermitizing" as is standard), which is noted $\gamma ^{\natural 0}$ in Ref. \cite{A50}. Whereas, $\gamma ^0$ is the ``$i=0$" matrix of the {\it field of Dirac matrices in the curved spacetime,}  $\gamma ^i\equiv e^i_{\ \,\alpha }\,\gamma ^\alpha $ ($\gamma ^\mu \equiv a^\mu _{\ \,\alpha }\,\gamma^{\natural  \alpha} $ in the notation of Refs. \cite{A47,A50}). Whether one mentions it or not, the field $\gamma ^i$ can be defined as soon as one has a tetrad field and a set of constant Dirac matrices $\gamma ^\alpha $ valid for the Minkowski metric, and it depends on both. In fact, the field $\gamma ^i$, or at least the field of the $\gamma ^0$ matrix, is needed to define the scalar product --- as shown precisely by Eq. (\ref{scalar product})$_1$. The field $\gamma ^i$ is also there in the Dirac equation, though not explicitly with the form (2.7) used by OST: (2.7) rewrites using (2.8)$_1$ in the slightly more standard form  \be\label{Dirac-standard} (\mathrm{i}\hbar \gamma ^i D_i-mc)\Psi =0. \ee
}
\be \label{scalar product}
(\Psi  \mid \Phi  ) \equiv \int \Psi^\dagger \sqrt{-g}\,\gamma ^{\widehat 0} \gamma ^0\, \Phi\ \dd^ 3{\bf x} \longrightarrow (\psi \, \widetilde{\mid} \,\phi  )= \int \psi^\dagger \phi\ \dd^ 3{\bf x} ,
\ee
is ensured by Eq. (\ref{a^0_0}) above, due to Eq. (19) in Ref. \cite{A50}. Recall that the scalar product has to be specified before one can state that some operator is Hermitian. In this Comment, {\it Hermitian} operators are stated to be so w.r.t. the ``flat" product (\ref{scalar product})$_2$.\\

Thus, starting from one tetrad field $(u_\alpha )$ in the Schwinger gauge (2.2)--(2.3) and getting then, in general, a non-Hermitian Hamiltonian [w.r.t. the product (\ref{scalar product})$_1$], the Hermitian Hamiltonian $\mathcal{H}$ obtained by OST using the transformation (2.14) [Eq. (\ref{T=a^{-1}I}) here] is just the one denoted $\mathrm{H}_\eta$ in Refs. \cite{A50, GorbatenkoNeznamov2011}, with here $\eta = (\sqrt{-g}\,e^0_{\ \,\widehat 0})^\frac{1}{2} {\bf 1}_4$. Now, consider another tetrad field $(u'_\alpha )$ in the Schwinger gauge (2.2)--(2.3). It is thus related with the first one $(u_\alpha )$ by the local Lorentz transformation $\Lambda $, Eq. (\ref{Lambda}): essentially, $\Lambda $ is the arbitrary rotation field $P(t,x)=(L^{\widehat a}{}_{\widehat c})^{-1}\in SO(3)$. Define $S'$, the local similarity transformation got by ``lifting" the local Lorentz transformation $\Lambda $ to the spin group. At this point, we could repeat {\it verbatim} what is written in Ref. \cite{A50} after the first sentence following Eq. (19). Thus, let $'\mathcal{H}\equiv \mathrm{H}_{\eta'}$ be the Hermitian Hamiltonian obtained by OST using the transformation (\ref{T=a^{-1}I}), but starting from the Schwinger tetrad field $(u'_\alpha )$ instead of the other one $(u_\alpha )$. (The notation $\mathcal{H}'$ designates something else in Ref. \cite{OST2013}.) The change from the Hamiltonian $\mathcal{H}=\mathrm{H}_\eta$ to the Hamiltonian $'\mathcal{H}=\mathrm{H}_{\eta'}$ is through the local similarity transformation $U=S'S^{-1}$, Eq. (20) of Ref. \cite{A50}. (Here, $U=S'$, because we have $S={\bf 1}_4$ as noted after Eq. (\ref{T=a^{-1}I}) above.) This implies that the Schr\"odinger equation $\mathrm{i}\hbar\frac{\partial \psi} {\partial t} = {\mathcal H}\psi$ is equivalent to  $\mathrm{i}\hbar\frac{\partial \psi'} {\partial t}=\, '\mathcal{H}\,\psi'$, with $\psi'=U^{-1}\psi$. Moreover, the similarity matrix $U$ is a unitary matrix. \{This property of the gauge transformations internal to the Schwinger gauge, derived in \cite{A50} from the invariance of the scalar product (\ref{scalar product})$_2$ under $U$, can be seen also from the fact that the Lorentz transformation (\ref{Lambda}) is a rotation.\} Hence the transformation
\be\label{psi'}
\psi \mapsto \psi '=U^{-1}\psi
\ee
is a unitary transformation internal to the Hilbert space $\mathsf{H}$ made, in view of (\ref{scalar product})$_2$, of the usual square-integrable functions of the spatial coordinates, $x\mapsto \psi (x)\in \mathbb{C}^4$ such that $(\psi \, \widetilde{\mid}\, \psi  )= \int \psi^\dagger \psi\ \dd^ 3{\bf x} <\infty $. {\it And\ } $'\mathcal{H}$ is physically inequivalent to $\mathcal{H}$, unless $\partial _t U=0$, i.e., unless the arbitrary rotation field $P$ is chosen independent of $t$. To be more precise, we have in fact \cite{A50,GorbatenkoNeznamov2011}:
\be\label{H' vs H}
'\mathcal{H}=U^{-1} \mathcal{H} U - i\hbar U^{-1} \partial _t U
\ee
(with here $U^{-1}=U^\dagger $, $U$ being a unitary matrix). 
\footnote{\ 
It is noted in Refs. \cite{A50, GorbatenkoNeznamov2011b} that $\mathcal{H}$ [respectively $'\mathcal{H}$] is also equal to the energy operator (the Hermitian part of the Hamiltonian) corresponding to the Schwinger tetrad $(u_\alpha )$ [respectively $(u'_\alpha )$]. Accounting for Eq. (\ref{a^0_0}) and for the fact that here $U$ is unitary, the same relation (\ref{H' vs H}) is got by using the general relationship \cite{A43} between two energy operators related by an admissible local similarity transformation. 
} 
This is actually the relation between two Hamiltonians exchanging by the most general ``operator gauge transformation" \cite{FoldyWouthuysen1950,Goldman1977} (in the present case a local similarity transformation, i.e. $U=U(X)$ is a regular complex matrix depending smoothly on the spacetime position $X$).\\

\paragraph{3. Physical inequivalence of $\mathcal{H}$ and $ '\mathcal{H}$. ---}\label{H vs H'} Indeed, contrary to what is stated by Gorbatenko \& Neznamov \cite{GorbatenkoNeznamov2013}, this inequivalence is what is expressed by Eq. (\ref{H' vs H}), unless $\partial _t U=0$. This had been already demonstrated in detail in Ref. \cite{A43}. It has been redemonstrated in Ref. \cite{A50}; that time with emphasis on the notions of a unitary transformation and of the mean value of a quantum-mechanical operator, invoked in Ref. \cite{GorbatenkoNeznamov2013}. Recall that the mean value $\langle \mathcal{H} \rangle $ of an operator such as the Hamiltonian operator $\mathcal{H}$ depends on the {\it state} $\psi $, belonging to the ``domain" (of definition) $\mathcal{D}$ of the operator $\mathcal{H}$. Here, in view of (\ref{scalar product})$_2$: 
\be\label{Def mean value}
\langle \mathcal{H} \rangle =\langle \mathcal{H} \rangle _\psi \equiv (\psi \, \widetilde{\mid}\, \mathcal{H}\psi  )= \int \psi^\dagger (\mathcal{H}\psi)\ \dd^ 3{\bf x} \qquad \mathrm{when}\ \psi \in \mathcal{D}.
\ee
(The domain $\mathcal{D}$ is a linear subspace of the whole Hilbert space $\mathsf{H}$, and $\mathcal{D}$ should be dense in $\mathsf{H}$. The precise definition of $\mathcal{D}$ should ensure that the integral above makes sense, for any $\psi \in \mathcal{D}$.) The mean value $\langle '\mathcal{H} \rangle$ for the corresponding state $\psi '$ after the transformation (\ref{psi'}) is given by the same Eq. (\ref{Def mean value}), with primes. It has been proved in Ref. \cite{A50} that, if the similarity matrix $U(t,x)$ in Eq. (\ref{H' vs H}) depends indeed on $t$, then not only the mean values $\langle \mathcal{H} \rangle$ and $\langle '\mathcal{H} \rangle$ are in general different, but in addition the difference $\langle '\mathcal{H} \rangle - \langle \mathcal{H} \rangle$ depends on the state $\psi \in \mathcal{D}$, so that the two Hamiltonians $\mathcal{H}$ and $'\mathcal{H}$ are physically inequivalent. \{This is true also \cite{A50} for the case of general ``non-Schwinger" gauge transformations, for which the regular complex matrix $U$ in Eq. (\ref{psi'}) is not unitary: then, the transformation (\ref{psi'}) is a unitary transformation between {\it two} Hilbert spaces \cite{A50,A43}.\} Moreover, the difference $\langle '\mathcal{H} \rangle - \langle \mathcal{H} \rangle$ can be calculated explicitly when the tetrad $(u_\alpha(t,x))$ is deduced from $(u'_\alpha (t,x))$ by the rotation of angle $\omega t$ around the vector $u_3(t,x)=u'_3(t,x)$. In that case, one has the explicit expression 
\be\label{U=S(Lambda)}
U(t)=e^{\omega tN},\qquad N\equiv (\alpha^1\alpha^2)/2\quad  (N^\dagger=-N).
\ee
Here, $\alpha^a \equiv  \gamma^{\widehat 0} \gamma^a$ ($a= 1,2,3$), in the notation of \cite{OST2013}, that is $\alpha '^j \equiv \gamma '^0\gamma '^j\ $ in the notation of Ref. \cite{GorbatenkoNeznamov2013}. Equation (\ref{U=S(Lambda)}) \cite{GorbatenkoNeznamov2013} applies to this more general situation as well. Indeed, the spin transformation $U$ (noted $R$ in Ref. \cite{GorbatenkoNeznamov2013}) that lifts the local Lorentz transformation (\ref{Lambda}) with $P^T$ the rotation of angle $\omega t$ around $u_3$ is independent of whether or not $u_3$ and $\partial _3$ coincide (as was the case in Ref. \cite{GorbatenkoNeznamov2013}). We get from (\ref{U=S(Lambda)})$_2$: $N=\frac{i }{2}\Sigma ^3\equiv \frac{i }{2}\mathrm{diag}(1,-1,1,-1)$ \cite{A50}, whence by (\ref{H' vs H}), (\ref{Def mean value}) and (\ref{U=S(Lambda)})$_1$ \cite{A50}:
\be\label{delta <H>}
\langle '\mathcal{H} \rangle - \langle \mathcal{H} \rangle =\frac{\omega}{2}\langle \Sigma ^3 \rangle
=\frac{\omega}{2}\int \left(\abs{\psi^0}^2 +\abs{\psi^2}^2-\abs{\psi^1}^2-\abs{\psi^3}^2\right)\,\dd^3{\bf x}.
\ee
This even holds true in the presence of an electromagnetic field \cite{A49}. This equation applies to {\it any} possible state $\psi \in \mathcal{D}$. It implies that, for the states $\psi\in \mathcal{D}$ such that $ \langle \Sigma ^3 \rangle \ne 0$, i.e. the integral in (\ref{delta <H>}) does not vanish, then definitely $\langle '\mathcal{H} \rangle \ne \langle \mathcal{H} \rangle$. The difference, $\delta =\frac{\omega }{2} \langle \Sigma ^3 \rangle$, depends on the state $\psi $ . For a normed state: $(\psi \widetilde{\mid }\psi )=1$, $\delta $ can take any value between $\frac{\omega }{2}$ and $-\frac{\omega }{2}$.  Note that $\omega $ is the rotation rate of the tetrad $(u_\alpha)$ w.r.t. $(u'_\alpha )$ and can be made arbitrarily large. Of course there are states for which $ \langle \Sigma ^3 \rangle=0$; e.g., when the metric (2.1) is the Minkowski metric of a flat spacetime, the coordinates $(x^i)$ being Cartesian and $u_3$ being parallel to $Ox^3$: an average state $\psi _\mathrm{av}\equiv (\psi _1+\psi _2)/\surd 2$, with $\psi _j\in  \mathcal{D}$ of the form $\varphi(x)A_j\ (j=1,2)$ where $\varphi(x)$ is a square-integrable scalar function and $A_j\in \mathbb{R}^4$ are the amplitude vectors of two plane wave solutions of the free Dirac equation, {\it with momentum parallel to $Ox^3$ i.e. to the rotation axis of the tetrad $(u_\alpha)$ w.r.t. $(u'_\alpha )$,} and with opposite helicities $\pm \frac{1}{2}$ (see Ref. \cite{Schulten2000}). (Such an average state has to be defined before the mean value is calculated, for the mean value is not linear.) In that particular metric as well as in a general one, the states for which $ \langle \Sigma ^3 \rangle=0$ are a very small subset of all physical states $\psi \in \mathcal{D}$. Moreover, by choosing another rotation field $P$, one may easily show that, also for the states having $ \langle \Sigma ^3 \rangle=0$, the energy mean values are indeterminate.

\paragraph{4. Inequivalence of $\mathcal{H}$ and $\mathcal{H}_\mathrm{FW}$ in the time-dependent case. ---}\label{FW} It has just been proved that, when the rotation field $P$ in Eq. (\ref{Lambda}) depends on $t$, the two Hermitian Hamiltonians (2.15) $\mathcal{H}$ and $'\mathcal{H}$ are inequivalent. This inequivalence would transmit automatically to that of the corresponding Foldy-Wouthuysen (FW) Hamiltonians (3.11), say $\mathcal{H}_\mathrm{FW}$ and $'\mathcal{H}_\mathrm{FW}$ --- if the FW transformation (3.2) would lead to an equivalent Hamiltonian to the starting one, i.e., if $\mathcal{H}$ were equivalent to $\mathcal{H}_\mathrm{FW}$, and $'\mathcal{H}$ to $'\mathcal{H}_\mathrm{FW}$. However, the issue of inequivalence enters the scene an {\it additional time} here, because the unitary transformation (3.2) has just the form (\ref{psi'})--(\ref{H' vs H}) (albeit with $U\rightarrow U^{-1}$), which, for the case $\partial _t U \ne 0$, is responsible for an inequivalence of the Hamiltonians before and after transformation. This issue may have been noted by Eriksen \cite{Eriksen1958}, perhaps even by Foldy \& Wouthuysen themselves \cite{FoldyWouthuysen1950,Goldman1977}. Eriksen \cite{Eriksen1958} limited the application of the FW transformation to ``non-explicitly time-dependent" transformations, i.e. indeed $\partial _t U = 0$ in Eqs. (\ref{psi'})--(\ref{H' vs H}). That issue has been discussed in detail by Goldman \cite{Goldman1977}, for the FW transformation applied to the Dirac equation in a flat spacetime in Cartesian coordinates in the presence of an electromagnetic field. He noted explicitly that, in the time-dependent case, the transformed Hamiltonian [here $'\mathcal{H}$ in (\ref{H' vs H})] ``is unphysical (in the sense of EEV's)" (energy expectation values), given that (in his case) ``the EEV's of the original $\mathcal{H}$ [are supposed to] have physical meaning". \\

The FW transformation $U$ (3.3) used in effect by OST has the form $U=\mathcal{N}.(\mathcal{N}^2)^{-1/2}\beta $ with $\mathcal{N}=\mathcal{N}(t)\equiv \beta \epsilon +\beta \mathcal{M}-\mathcal{O}$ a time-dependent operator (implicitly assumed to be Hermitian and have an inverse), where ${\cal H}=\beta {\cal M}+{\cal E}+{\cal O}$ is a decomposition, stated by OST, of the Hermitian Hamiltonian $\mathcal{H}$, Eq. (3.1);
\footnote{\
Any operator acting on four-component functions decomposes uniquely into ``even" and ``odd" parts, i.e. commuting and anticommuting with $\beta \equiv \gamma ^{\widehat 0}\equiv \mathrm{diag}(1,1,-1,-1)$ \cite{FoldyWouthuysen1950,Eriksen1958}. The decomposition (3.1) of the given operator ${\cal H}$ is hence unique for {\it any} even operator $\cal{M}$ which is also {\it given}. In the present case, $\cal{M}$ appears to be given by Eq. (3.8). Then one can easily express ${\cal E}$ and ${\cal O}$ in terms of the Hermitian operators ${\cal H}$, $\beta $ and ${\cal M}$, and check that $\mathcal{N}$ is indeed Hermitian.
}
and $\epsilon \equiv \sqrt{{\cal M}^2+{\cal O}^2}$. Assume for simplicity that there is a subdomain $\mathcal{D}_\mathrm{e}(t) \subset \mathcal{D}$ to which the restriction of $\mathcal{N}(t)$ has a  decomposition in eigenspaces:  $\mathcal{D}_\mathrm{e}=\oplus _j \mathrm{E}_j$ and $\mathcal{N}_{\mid \mathcal{D}_\mathrm{e}}=\Sigma _j \lambda _j\, \mathrm{Pr}_{\mathrm{E}_j}$ with $\lambda _j(t)$ nonzero reals and $\mathrm{Pr}_{\mathrm{E}_j}$ the orthogonal projection on the eigenspace $\mathrm{E}_j(t)$. Then, if $\psi \in \mathrm{E}_j$, we have $\mathcal{N}.(\mathcal{N}^2)^{-1/2}\psi =\varepsilon _j\,\psi $ with $\varepsilon _j\equiv \mathrm{sgn}(\lambda _j)=\pm1$. Hence, defining $\mathcal{D}_\mathrm{e}^+=\oplus _{\varepsilon _j=+1} \mathrm{E}_j$ and the like for $\mathcal{D}_\mathrm{e}^-$, we have $\mathcal{D}_\mathrm{e}=\mathcal{D}_\mathrm{e}^+\oplus \mathcal{D}_\mathrm{e}^-$ and $[\mathcal{N}.(\mathcal{N}^2)^{-1/2}]_{\mid \mathcal{D}_\mathrm{e}}=\mathrm{Pr}_{\mathcal{D}_e^+} -\mathrm{Pr}_{\mathcal{D}_e^-}$, whence $U_{\mid \beta ^{-1}(\mathcal{D}_\mathrm{e}^\pm)}=\pm \beta $ and 
\be\label{U Proj}
U_{\mid \beta ^{-1}(\mathcal{D}_\mathrm{e})}=\left(\mathrm{Pr}_{\mathcal{D}_e^+ }-\mathrm{Pr}_{\mathcal{D}_e^-} \right)\beta ,
\ee
with in fact $\beta ^{-1}=\beta $. Now, clearly the eigenspaces $\mathrm{E}_j$ of the operator $\mathcal{N}(t)$ evolve with time in a general metric, due to the time-dependence of the metric and that of the tetrad, hence $\mathcal{D}_\mathrm{e}^+$ and $\mathcal{D}_\mathrm{e}^-$ also evolve. So Eq. (\ref{U Proj}) confirms that $U$ is time-dependent. Thus in the general case the FW transformation (3.3) depends on time, so that the FW Hamiltonian (3.11) is physically inequivalent to the starting Hermitian Hamiltonian $\mathcal{H}$ (2.15). This is in addition to the physical inequivalence of two Hermitian Hamiltonians $\mathcal{H}$ and $'\mathcal{H}$ (2.15) got from two choices of the Schwinger tetrad, proved at points  \hyperref[NU-H]{({\bf 2})} and \hyperref[H vs H']{({\bf 3})}, and which applies even in the case of a Minkowski spacetime in Cartesian coordinates.

\paragraph{5. Semi-classical limit and comparison with classical spinning particles. ---} The quantum equation of motion of spin (3.15): $\frac{d\Mat{ \Pi}}{dt}=\frac{i}{\hbar}[{\mathcal H}_\mathrm{FW},\Mat{ \Pi}]$, should be ambiguous as is $\mathcal{H}_\mathrm{FW}$. In fact, the explicit dependence of $\mathcal{F}^a_{\ \,b}$ on the choice of the Schwinger tetrad field has been noted in Eq. (\ref{F'}). Hence, the three-vector operators of the angular velocity of spin precessing $\Mat{\Omega}_{(1)}$ and $\Mat{\Omega}_{(2)}$, Eqs. (3.16) and (3.17), depend {\it a priori} also on that choice. This dependence may survive in the semi-classical limit (3.19)--(3.20). To check these two points would be somewhat cumbersome. However, we note that the operator $p_a\equiv  -i\hbar \frac{\partial }{\partial x^a}$, as well as the c-number $p_a$ given by (4.30), are independent of the choice. Therefore, using $PP^T={\bf 1}_3$, one finds that $\epsilon '$ in Eq. (3.21) is invariant under the change of the Schwinger tetrad involving the substitution (\ref{F'}), for
\be\label{F=F'}
\delta^{cd}{\mathcal F}'^a{}_c\,{\mathcal F}'^b{}_d
\,p_a\,p_b = \delta^{cd}{\mathcal F}^a{}_e\,P^e_{\ \,c}\,{\mathcal F}^b{}_f\,P^f_{\ \,d}\,p_a\,p_b = \delta^{ef}{\mathcal F}^a{}_e\,{\mathcal F}^b{}_f\,p_a\,p_b.
\ee
This applies whether in (3.21) one considers $p_a$ and $\epsilon '$ as operators or as c-numbers. It follows, using again (\ref{F=F'}), that the semi-classical velocity operator $\frac{dx^a}{dt}$ as given by the last member of Eq. (3.23) is invariant under the change of the Schwinger tetrad. The semi-classical velocity operator in the Schwinger frame (3.24) is covariant under the change of the Schwinger tetrad: $v'_a=P^b_{\ \,a}\, v_b$, if $\epsilon '$ is regarded as a c-number in (3.24).\\

The degree to which ``the classical equation of the spin motion (4.22) agrees with the quantum equation (3.15) and with the semiclassical one (3.18)" \cite{OST2013} does not appear very clearly: these are three rather complex expressions which do not seem to coincide, and as noted above the quantum equation (3.15) looks ambiguous, to the very least {\it a priori}. We note that the classical spin rate $\Mat{\Omega }$ (4.36), as well as [using (\ref{F=F'}) with $p_a \rightarrow \pi_a$] the classical Hamiltonian (4.38), are independent of the choice of the Schwinger tetrad. In their conclusion, the authors of Ref. \cite{OST2013} state that a ``complete consistency of the quantum-mechanical and classical descriptions of spinning particles is also established using the Hamiltonian approach in Sec. IV B". However, the quantum-mechanical description --- in particular, the Hermitian Hamiltonian operator, be it $\mathcal{H}$ (2.15) or $\mathcal{H}_\mathrm{FW}$ (3.11) --- is seriously non-unique as demonstrated at points \hyperref[NU-H]{({\bf 2})}, \hyperref[H vs H']{({\bf 3})} and \hyperref[FW]{({\bf 4})} above. Whereas, the classical description is unique as we just saw. The ambiguity of the covariant Dirac theory regards the energy mean values and eigenvalues \cite{A50,A43}, but the probability current and its motion are unambiguous. In the wave packet approximation, the latter motion can be rewritten as the exact equations of motion of classical (non-spinning) particles in the electromagnetic field, without any use of the semi-classical limit $\hbar\rightarrow 0$ \cite{A46}.

% \newpage

\end{document}